\documentclass[twocolumn,prl,amsmath,amssymb,showpacs,superscriptaddress]{revtex4-1}
\usepackage{epsf}      
\usepackage{graphicx}
\usepackage{color}
\usepackage{gensymb}
\usepackage{sidecap}

\begin{document}
\title {Neutron diffraction and magnetic properties of Co$_2$Cr$_{1-x}$Ti$_x$Al Heusler alloys}

\author{Priyanka Nehla}
\affiliation{Department of Physics, Indian Institute of Technology Delhi, Hauz Khas, New Delhi-110016, India}
\author{Yousef Kareri}
\affiliation{School of Physics, The University of New South Wales, Kensington, 2052 NSW, Sydney, Australia}
\author{Guru Dutt Gupt}
\affiliation{Department of Physics, Indian Institute of Technology Delhi, Hauz Khas, New Delhi-110016, India}
\author{James Hester}
\affiliation{Australian Centre for Neutron Scattering, Australian Nuclear Science and Technology Organisation (ANSTO), New Illawarra Road, Lucas Heights NSW 2234, Australia }
\author{P. D. Babu}
\affiliation{UGC-DAE Consortium for Scientific Research, Trombay, Mumbai-400085 India.}
\author{Clemens Ulrich}
\affiliation{School of Physics, The University of New South Wales, Kensington, 2052 NSW, Sydney, Australia}
\author{R. S. Dhaka}
\email{rsdhaka@physics.iitd.ac.in}
\affiliation{Department of Physics, Indian Institute of Technology Delhi, Hauz Khas, New Delhi-110016, India}

\date{\today}

\begin{abstract}

We report the structural, magnetic, and magnetocaloric properties of Co$_2$Cr$_{1-x}$Ti$_x$Al ($x=$ 0--0.5) Heusler alloys for spintronic and magnetic refrigerator applications. Room temperature X-ray diffraction and neutron diffraction patterns along with Rietveld refinements confirm that the samples are of single phase and possess a cubic structure. Interestingly, magnetic susceptibly measurements indicate a second order phase transition from paramagnetic to ferromagnetic where the Curie temperature (T$_{\rm C}$) of Co$_2$CrAl increases from 330~K to 445~K with Ti substitution. Neutron powder diffraction data of the $x=$ 0 sample across the magnetic phase transition taken in a large temperature range confirm the structural stability and exclude the possibility of antiferromagnetic ordering. The saturation magnetization of the $x=$ 0 sample is found to be 8000~emu/mol (1.45~$\mu_{\rm B}$/{\it f.u.}) at 5~K, which is in good agreement with the value (1.35$\pm$0.05~$\mu_{\rm B}$/{\it f.u.}) obtained from the Rietveld analysis of the neutron powder diffraction pattern measured at temperature of 4~K. By analysing the temperature dependence of the neutron data of the $x=$ 0 sample, we find that the change in the intensity of the most intense Bragg peak (220) is consistent with the magnetization behavior with temperature. Furthermore, an enhancement of change in the magnetic entropy and relative cooling power values has been observed for the $x=$ 0.25 sample. Interestingly, the critical behavior analysis across the second order magnetic phase transition and extracted exponents ($\beta\approx$ 0.496, $\gamma\approx$ 1.348, and $\delta\approx$ 3.71 for the $x=$ 0.25 sample) suggest the presence of long-range ordering, which deviates towards 3D Heisenberg type interactions above T$_{\rm C}$, consistent with the interaction range value $\sigma$.

\end{abstract}
 
\maketitle

\section{\noindent ~Introduction}

The Heusler alloys are intermetallic compounds with the general formula X$_2$YZ for full and XYZ for half Heusler compounds, where X and Y are transition metal elements and Z represents an $sp$ element like Al, Ga, Si, or another main group metal \cite{Felser16}. In recent years, the Co$_2$Cr$_{1-x}$T$_x$Z (T = transition metal) Heusler alloys are considered potential candidates for various practical applications, as these materials have theoretically been predicated to be half metallic ferromagnets (HMF) with 100\% spin polarization \cite{GalanakisPRB02, ZhangJMMM04, MiuraJAP06}, and possess excellent structural stability, low magnetic damping \cite{HusainSR16}, anomalous Hall conductivity \cite{HusmannPRB06}, large negative magnetoresistance \cite{BlockJSSC03} as well as a high Curie temperature (T$_{\rm C}$) above room temperature \cite{WurmehlAPL06,KublerPRB07}. These are most favorable properties for the technological development of spintronic devices operating at and above room temperature \cite{Felser16}. However, understanding the role of temperature/field dependence of the magnetic properties and local atomic ordering \cite{SasiogluPRB05, PriyankaJALCOM19, RaniPRA18, KublerPRB07} is crucial in making real use of these materials in device applications \cite{PalmstromPCGCM16}. In general, the Co$_2$-based Heusler alloys possess the fully ordered L2$_1$ structure and the experimentally determined magnetic moment is in agreement with the theoretical calculated value using the Slater-Pauling rule \cite{UmetsuPRB05,KublerPRB07,FecherJAP06}. Disorder can transform the L2$_1$ structure to B2 type (disorder between Y and Z atoms) or A2 type (disorder between X, Y and Z atoms). In case of the Co$_2$CrAl alloy, the increasing Co-Cr type disorder \cite{ZagrebinPBCM2017, KudryavtsevPRB08, RaniPRA18} can drastically reduce the spin polarization and therefore the total magnetic moment due to an antiferromagnetic coupling of the antisite Cr with the first nearest-neighbor ordinary-site Cr and Co atoms \cite{LiSSC10, MiuraPRB04}. With this respect, using powder neutron diffraction technique Mukadam {\it et al.} quantified the site disorder and studied its effect on the spin polarization of the NiFeMnSn Heusler alloy, which is predicted to show nearly half-metallic nature in band structure calculations \cite{MukadamPRB16}. Therefore, the role of disorder between X, Y and Z atoms is of crucial importance in order to control the magnetic moment, the half-metallicity and the spin polarization in Heusler alloys \cite{ZagrebinPBCM2017, KudryavtsevPRB08, MiuraPRB04, FengJMMM15}. 

On the other hand, Kandpal {\it et al.} showed that antisite disorder of 10--12\% between Co/Ti  does not change the half-metallic character and magnetization in Co$_2$TiSn \cite{KandpalJPD07}. Moreover, first principle calculations demonstrated that Co$_2$TiZ alloys are half-metallic and ferromagnetic in nature even when the Z element is from the III, IV or V group. However, it was suggested that disorder between Ti/Al reduces the half-metallicity \cite{LeeJAP05}. One of the most favorable properties of these Heusler alloys for potential future applications is the fact that they offer a direct tunability of their electronic and magnetic properties \cite{Feng17, PriyankaXiv19, MizusakiJAP12} through the substitution of one of the transition metals with another element. Interestingly, the magnetic properties of Co$_2$Cr$_{1-x}$Ti$_x$Al Heusler alloys show a decrease in the total magnetization and the sample with $x=$ 0.5 possess the most stable half metallicity \cite{Feng17}. Y. Feng, and X. Xu reported that with increasing Ti concentration the Fermi level shifts from the bottom to top of the minority spin gap and is located in the middle of the gap for a Ti concentration of $x=$ 0.5 \cite{Feng17}. Therefore, the effect of substitution in the ternary Heusler alloys opens the gate to discover new materials \cite{Graf09, HusmannPRB06, DiMasi93} and hence the understanding of the physical properties is vital for their practical applications.

In recent years tremendous research interest has also been generated in the Heusler alloys due to their magnetocaloric properties, which offer a large potential for  applications in magnetic refrigerators \cite{SinghAM16,PandaJALCOM15}. It is important to note that magnetic refrigeration has advantages over compression/expansion techniques due to their low cost, highly efficient and most important environment friendly operation since no compressor/refrigerator gas is required \cite{FrancoEPL2007}. Therefore, the scientific interest in materials with magnetic cooling near room temperature has been grown significantly. For example, compounds with the rare earth element Gd have a large magnetocaloric effect (MCE) and have been actively investigated for the use in magnetic refrigerator \cite{FrancoEPL2007}. In particular,  intermetallic compounds such as Gd-Si-Ge, Mn-Fe-P-Si, and La-Fe-Si \cite{Bruck07} were found to be highly efficient. However, at the same time rare earth materials are very expensive. An alternative to the rare earth compounds, the transition metal based magnetocaloric materials, in particular Heusler alloys are one of the suitable candidates for magnetic refrigeration near room temperature because of their large MCE and the adiabatic temperature change \cite{Hu00}. Of particular interest are Ni-based alloys since they show a first order magnetic phase transition and a large magnetic entropy change within a small temperature range \cite{Hu00}. However, under an applied magnetic field and temperature variation they exhibit large thermal and magnetic hysteresis, which limits their use for practical applications. A few Ni-Mn based materials have already been explored \cite{Hu00}, which show giant MCE, but they are not suitable for practical application because of their thermal and magnetic hysteresis which are detrimental to the refrigerant capacity \cite{HalderJAP11}. In this regard, a few studies have reported on second order magnetic phase transition materials which exhibit a large change in the magnetic entropy \cite{HalderJAP11, WangJAP09, KumarEPL18}. On the other hand, the Co-based Heusler alloys remain largely unexplored for this purpose \cite{PandaJALCOM15, PriyankaXiv19}. Therefore, it is desirable to search for new materials which exhibit large MCE and negligible thermal and magnetic hysteresis with a second order magnetic phase transition.

In this investigation we have studied the structural, magnetic, and magnetocaloric properties of the Co$_2$Cr$_{1-x}$Ti$_x$Al Heusler alloys with Ti-concentrations of $x=$ 0, 0.25, and 0.5. Using Rietveld refinements of powder X-ray diffraction (XRD) and powder neutron diffraction (ND) patterns we were able to confirm that the samples possess a stable single phase with a cubic crystal structure. Interestingly, we observe a second order magnetic phase transition from the paramagnetic (PM) to the ferromagnetic (FM) state where the Curie temperature can be tuned with Ti substitution from 330~K ($x=$ 0 sample) to 445~K for the $x=$ 0.5 sample. This property makes these materials extremely useful for technological applications. The saturation magnetization of the $x=$ 0 sample is found to be 8000~emu/mol (1.45~$\mu_{\rm B}$/{\it f.u.}), which is in good agreement to the value of 1.35$\pm$0.05~$\mu_{\rm B}$/{\it f.u.} extracted by the Rietveld analysis of ND pattern measured at 4~K. The analysis of the temperature dependent ND data of the $x=$ 0 sample shows that the change in the magnetic intensity of most intense Bragg peak (220) is consistent with the magnetization behavior with temperature. We found a conventional magnetocaloric effect across the PM-FM phase transition and the $x=$ 0.25 sample shows the highest $\Delta S_m$ and relative cooling power values at 9~Tesla. More interestingly, using critical behavior analysis across the second order magnetic phase transition for the $x=$ 0.25 sample, the obtained exponent $\beta\approx$ 0.496 suggests the presence of long-range ordering below T$_{\rm C}$, which deviates from mean field towards 3D Heisenberg type interactions above T$_{\rm C}$ where the exponents are $\gamma\approx$ 1.348, and $\delta\approx$ 3.71 \cite{RoyPRB19}.

\section{\noindent ~Experimental Details}

Polycrystalline Co$_2$Cr$_{1-x}$Ti$_x$Al ($x=$0, 0.25, 0.5) samples have been synthesized from high purity ($>$99.99\%) elements Co, Cr, Ti and Al from Alfa Aesar and/or Sigma Aldrich. An arc furnace from Centorr vacuum industries, USA was used to prepare an ingot by melting the stoichiometric amount of the starting materials on a water cooled Cu hearth in Argon atmosphere. In order to improve the homogeneity the ingot was flipped and melted 4--5 times. The loss in weight was less than 1\% after the melting process. Further, we wrapped the ingot in a Mo foil and sealed it in a quartz tube under vacuum for the final annealing at 575~K for 10 days. To determine the quality of the obtained samples, powder XRD measurements were performed at room temperature on a Panalytical diffractometer using the Cu K$\alpha$ ($\lambda$ = 1.5406~$\rm \AA$) radiation. Magnetization measurements were performed using a physical property measurement system (PPMS) of the company Quantum Design, USA. Neutron powder diffraction experiments have been performed using the high intensity neutron diffractometer WOMBAT \cite{Studer06} and the high resolution diffractometer ECHIDNA \cite{Liss06,Avdeev18} at the OPAL research reactor at ANSTO, Australia. At the high intensity diffractometer WOMBAT a wavelength of $\lambda=$ 1.633~\AA~was selected using a Ge(113) monochromator. Scans at various temperatures were recorded on heating from the base temperature of around 4~K up to a temperature well above the magnetic phase transition. In addition, diffraction patterns were collected at selected temperatures above and below the magnetic phase transition on the high resolution diffractometer ECHIDNA using a wavelength of $\lambda=$ 1.622~\AA~obtained by a Ge(335) monochromator. The measured diffraction patterns were analyzed with the Rietveld refinement method using the FullProf package \cite{Carvajal93}.

\section{\noindent ~Results and Discussion}

\begin{figure*}
\includegraphics[width=7.0in]{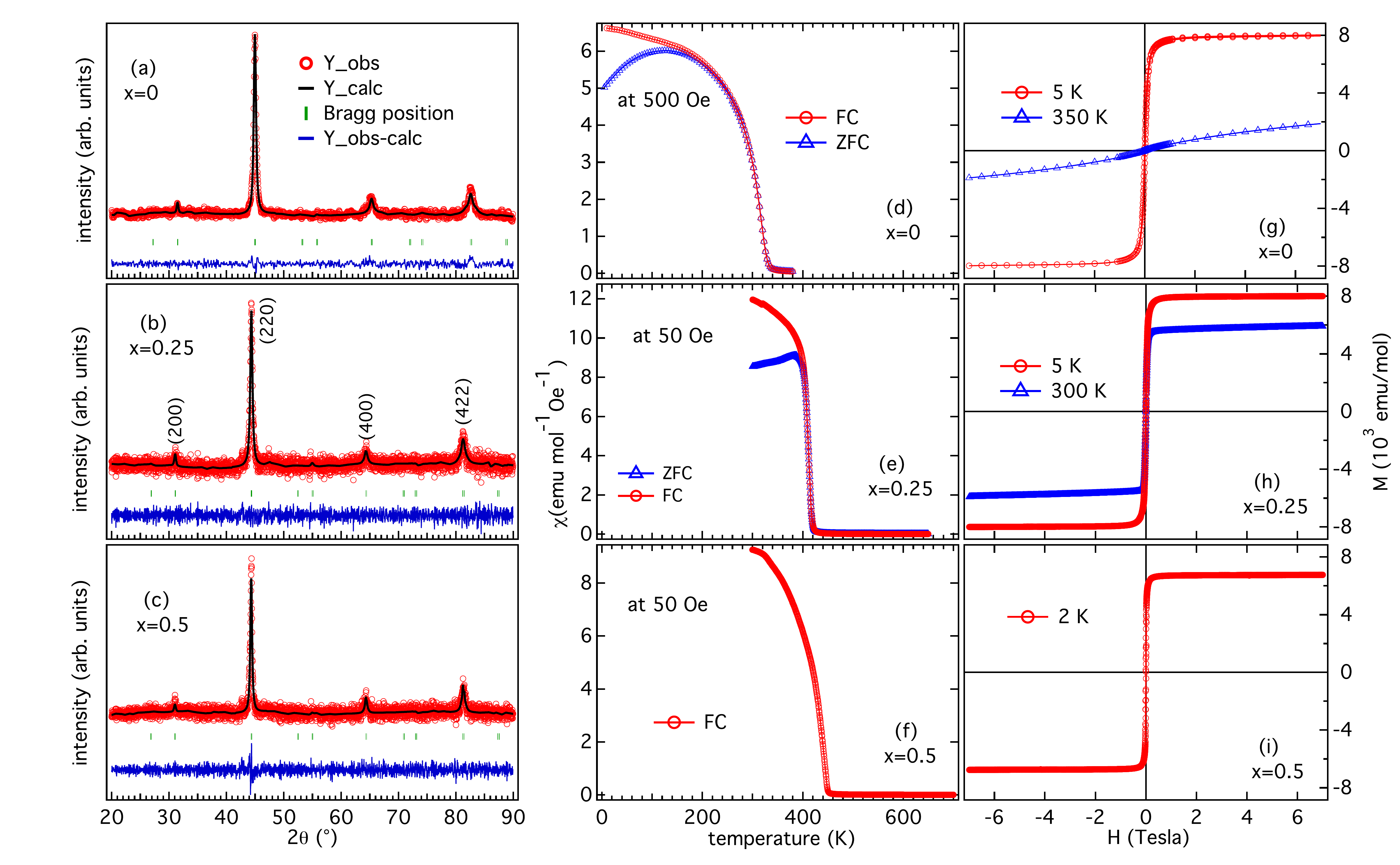}
\caption{(a--c) Powder XRD data (red open circles) along with the Rietveld refinement curves (black solid line). The blue solid line at the bottom shows the difference between the observed and calculated patterns, the green vertical tick marks are corresponds to the position of the Bragg reflections. (d--f) the variation of magnetic susceptibility ($\chi$) with temperature and  (g--i) isothermal magnetization M of Co$_2$Cr$_{1-x}$Ti$_x$Al ($x=$ 0, 0.25, 0.5) samples.}
\label{fig1}
\end{figure*}

Figures~\ref{fig1}(a--c) show the powder XRD patterns of the Co$_2$Cr$_{1-x}$Ti$_x$Al ($x=$ 0, 0.25, 0.5) samples measured at room temperature together with the Rietveld refinement profiles. We observe Bragg peaks correspond to  the (200), (220), (400) and (422) planes only, which indicates a cubic structure for all these samples. The absence of additional peaks confirms the phase purity of the samples and the absence of the (111) reflection indicates a B2-type ordering in these samples i.e. Cr and Al may partially exchange their atomic positions. Usually the presence of the (200) reflection confirms the ordering of the Co sublattice; however, due to the similar X-ray scattering cross-sections of Co (0.24) and Cr (0.27) \cite{SvyazhinJETP13}, no firm conclusion can be drawn about the local disorder between the two elements, which plays crucial role for the determination of the magnetic moment. On the other hand, theoretical calculations have predicted that the perfectly ordered L2$_1$ structure is more stable in Co$_2$CrAl \cite{ZagrebinPBCM2017}. We have performed Rietveld refinements of the XRD patterns using the F{\it m$\bar{3}$m} (no. 225) space group where the background has been fitted by linear interpolation between the experimental data points. The refinements provide reasonably good fits to the experimental data by assigning the Wyckoff positions (1/4, 1/4, 1/4), (0, 0, 0), and (1/2, 1/2, 1/2) to the Co, Cr/Ti and Al atoms, respectively. For the $x=$ 0 sample, the lattice parameter $a$ obtained from the refinement is found to be 5.742~\AA, which is consistent with the reported value 5.726~\AA~\cite{KudryavtsevPRB08}. For both Ti substituted samples ($x=$ 0.25, 0.5) the lattice parameter $a$ increases to $\approx$5.816~\AA. Note that the reported value of the lattice parameter $a$ for Co$_2$TiAl is 5.849~\AA~\cite{Graf09}. The change in the lattice parameter is consistent with the increase in the atomic radius from Cr (128~pm)  to Ti (147~pm), which results in the expected increase of the total unit cell volume ($a^3$) of Co$_2$Cr$_{1-x}$Ti$_x$Al with the corresponding Ti concentration at the Cr site. 

The magnetic susceptibility ($\chi$) measured as a function of temperature in zero-field-cooled (ZFC) and field-cooled (FC) protocol are shown in Figs.~\ref{fig1}(d--f) for Co$_2$Cr$_{1-x}$Ti$_x$Al ($x=$ 0--0.5) samples. For the $x=$ 0 sample the magnetic phase transition is at T$_{\rm C}$ $\approx$ 330~K, which is in good agreement to the value reported in ref.~\cite{HusmannPRB06}. An important result is the fact that T$_{\rm C}$ increases from 330~K to 410~K for the $x=$ 0.25 sample and further to 445~K for the $x=$ 0.5 sample. On the other hand, T$_{\rm C}$ of Co$_2$TiAl has been reported to be about 120~K \cite{DiMasi93}. The partial substitution of Cr with Ti enhances T$_{\rm C}$ significantly above room temperature, which make these materials extremely useful for practical applications in spintronics and magnetic refrigerators. The smooth behavior and the absence of a thermal hysteresis at the phase transition (data not shown here) from the paramagnetic to the ferromagnetic state are indications of a second order phase transition in these samples. Figures~\ref{fig1}(g--i) show the variation of magnetization with applied magnetic field at constant temperature, which indicates the soft ferromagnetic nature of these samples. There is no significant change in the total magnetization [8000~emu/mol (1.45~$\mu_{\rm B}$/{\it f.u.}) at 5~K] between the $x=$ 0 and 0.25 samples [Figs.~\ref{fig1}(g, h)]. The isothermal magnetization of the $x=$ 0.25 sample shows a ferromagnetic nature at 300~K. On the other hand, the magnetization of $x=$ 0.5 sample decreases to about 7000~emu/mol (1.25~$\mu_{\rm B}$/{\it f.u.}), see Fig.~\ref{fig1}(i). We note that in the present case the magnetic moment of the $x=$ 0 sample is significantly lower than the calculated value of 3~$\mu_{\rm B}$ as per the Slater-Pauling (SP) rule (given by M$_t=Z_t-$ 24, where M$_t$ is the total spin magnetic moment per f.u. in $\mu_{\rm B}$ and Z$_t$ is the total number of valence electrons \cite{GalanakisPRB02}), if we consider the sample in fully ordered state \cite{BlockPRB04}. On the other hand, the experimental value of the magnetic moment has been reported to be in the range of 1.5 to 3~$\mu_{\rm B}$/{\it f.u.} for Co$_2$CrAl \cite{ElmersPRB03, BlockPRB04, KudryavtsevPRB08}. Block {\it et al.} performed band structure calculations and reported that the magnetic moments are mainly localized at the Co ($\approx$0.8~$\mu_{\rm B}$) and Cr ($\approx$1.6~$\mu_{\rm B}$) sites. This indicates a ferromagnetic interaction between the transition metals (Co and Cr) and as a consequence the formation of the energy gap of about 0.2~eV in the minority bands \cite{BlockPRB04}. It has also been suggested that the experimentally observed lower value of the magnetic moment in Co$_2$CrAl can be influenced by the local structural anti-site disorder between Co and Cr, which might introduce a local antiferromagnetic ordering \cite{MiuraPRB04}. 

\begin{figure}
\includegraphics[width=3.25in]{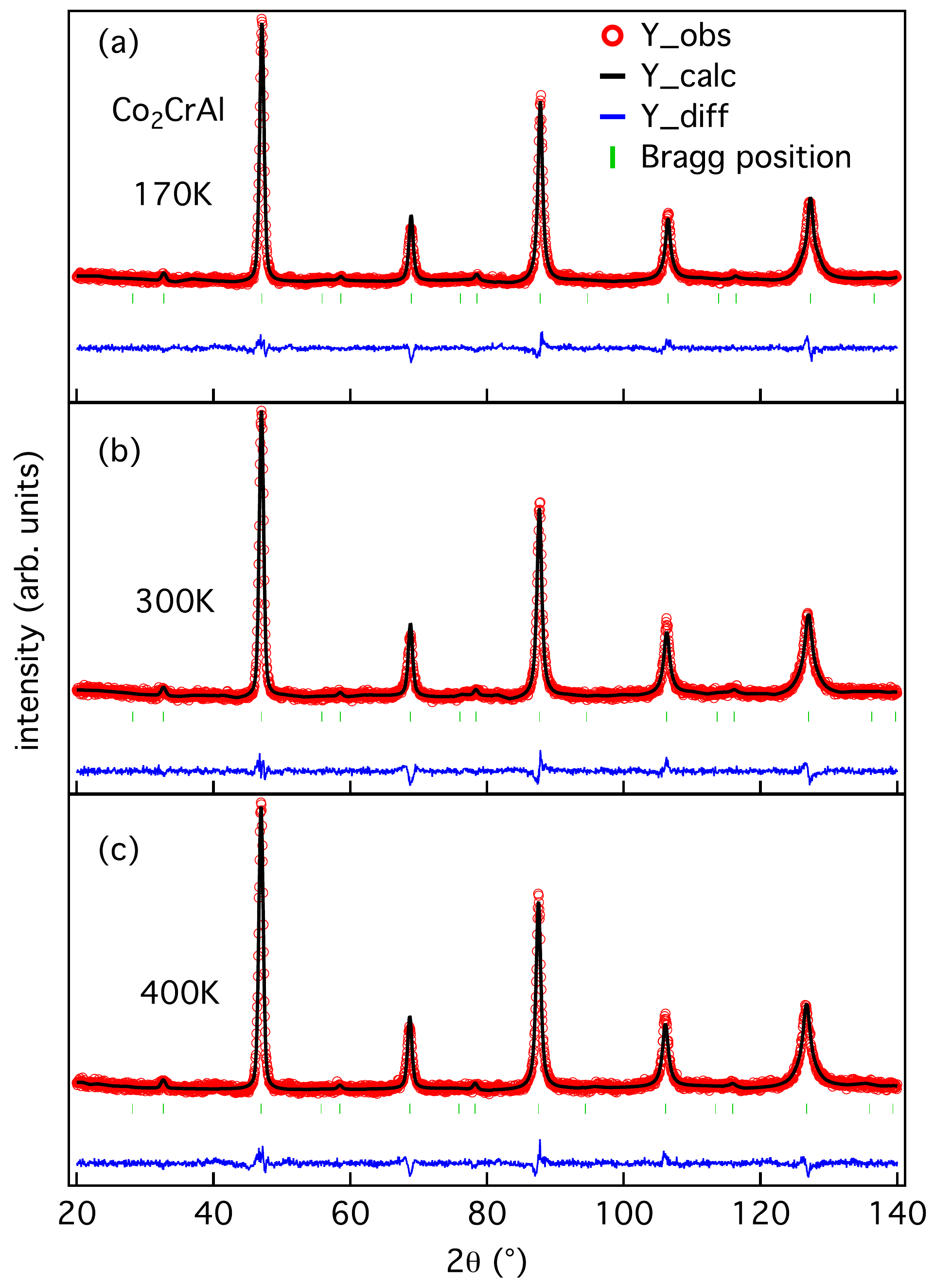}
\caption{High resolution neutron powder diffraction data of the $x=$ 0 sample (T$_{\rm C}$ = 330~K) measured at the instrument ECHIDNA ($\lambda=$ 1.622\AA) at temperatures of 170~K, 300~K and 400~K. The results of the Rietveld refinements are shown as black solid lines and the blue lines indicate the difference between the observed and calculated patterns. The vertical tick marks correspond to the structural Bragg peak positions.}
\label{fig2}
\end{figure}

The total number of valence electrons change from 27 in Co$_2$CrAl to 25 in Co$_2$TiAl. Therefore, the SP behavior predicts that the saturation magnetic moment should decrease with the partial substitution of Ti at the Cr site i.e. 2~$\mu_{\rm B}$ for Co$_2$Cr$_{0.5}$Ti$_{0.5}$Al and 1~$\mu_{\rm B}$ for Co$_2$TiAl. Recently, Mizusaki {\it et al.} used magnetic Compton scattering and reported a value of the moment of about 1.5~$\mu_{\rm B}$ and 1~$\mu_{\rm B}$ for Co$_2$CrAl and Co$_2$TiAl, respectively \cite{MizusakiJAP12}. The lower  value of the magnetic moment may be due to constant $sp-d$ magnetic interactions in Co$_2$CrAl \cite{MizusakiJAP12}. Also, the existence of an orbital moment of the Co ions has been suggested \cite{FelserJPCM03, GalanakisPRB05}. For example, an orbital moment of --0.15~$\mu_{\rm B}$/{\it f.u.} has been attributed to the Co ions in the Co$_2$TiAl Heusler alloy \cite{FelserJPCM03, GalanakisPRB05}. In the present case we observe a value of the moment of about 1.25~$\mu_{\rm B}$/{\it f.u.} for the $x=$ 0.5 sample, which is still lower than the SP behavior (2~$\mu_{\rm B}/${\it f.u.}). Interestingly, it has been reported that the Co--Cr(Ti) type disorder in Co$_2$CrAl alloys significantly reduces the spin polarization, but not the Cr(Ti)--Al type disorder \cite{MiuraPRB04}. Therefore, it becomes important to determine the type of disorder in this Heulser alloy. In general one can get an idea about the degree of disorder in X$_2$YZ Heusler alloys by taking the intensity ratio of (111) and (200) peaks, conventional X-ray diffraction can not give accurate results as nearby elements in the periodic table have almost same value of atomic scattering amplitude \cite{ZhuPRB17}. Note that in our case (Co$_2$CrAl), the X-ray scattering amplitudes of Co and Cr are nearly same, which makes it difficult to perform a quantitative analysis of the disorder using conventional X-ray diffraction. However, due to the different scattering amplitudes, neutron diffraction is more sensitive to the antisite disorder as compare to the X-ray diffraction. Also, neutron study can help if there is antiferromagnetic/ferrimagnetic alignment \cite{HalderPRB2011} of the Co and Cr spins, which may explain the lower experimental value of the magnetic moment in Co$_2$CrAl. For example, the neutron diffraction measurements shows a ferrimagnetic alignment between two different Mn atoms present in Mn$_2$VGa \cite{RameshJPCM12}. Therefore, it is vital to perform neutron diffraction measurements on the Co$_2$CrAl Heusler alloy across the magnetic phase transition in a large temperature range. 

\begin{figure}
\includegraphics[width=3.3in]{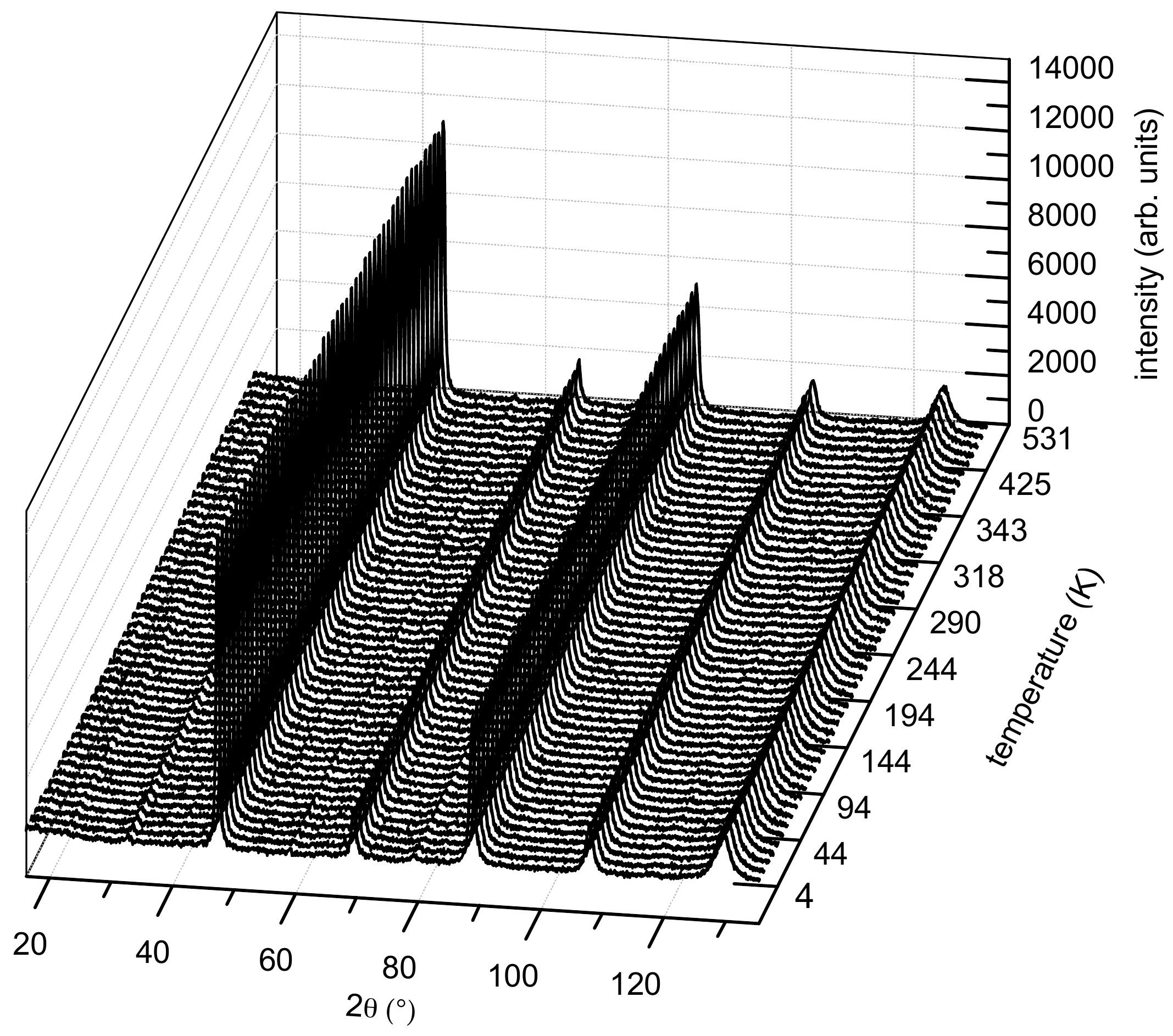}
\caption{The neutron powder diffraction patterns of Co$_2$CrAl as a function of temperature (4~K to 531~K) obtained with the high intensity instrument WOMBAT.}
\label{fig3}
\end{figure}

In order to understand the local atomic ordering and magnetic behavior, we have carried out the neutron powder diffraction measurements across the magnetic phase transition temperature of Co$_2$CrAl. We first used the high resolution diffractometer ECHIDNA and performed measurements at a sample temperature of 400~K, 300~K, and 170~K. The corresponding diffraction pattern is presented in Figs.~\ref{fig2} (a--c) along with the corresponding Rietveld refinements. The absence of the (111) Bragg reflection at all temperatures indicates the B2 structural order in the sample, which suggests for disorder between the Cr and Al atoms. As discussed before, the measured total magnetic moment of the $x=$ 0 sample (1.45~$\mu_{\rm B}$/{\it f.u.}) from the magnetization measurements is significantly lower than predicted from SP rule (3~$\mu_{\rm B}$). However, disorder between Cr and Al is not expected to decrease the magnetic moment of Co-based Heusler alloys \cite{MiuraPRB04}. Here the lower value of experimental magnetic moment can possibly be attributed to a certain type of disorder (between Co and Cr) in the system due to ferrimagnetic or antiferromagnetic spin alignment \cite{ZagrebinPBCM2017}. The Rietveld refinement of the neutron diffraction pattern taken at 400~K, i.e. above T$_{\rm C}$ [see Fig.~\ref{fig2}(c)], shows that the sample has a single cubic phase with the space group F{\it m$\bar{3}$m} and the refined lattice constant was found to be 5.708~\AA. In order to determine the disorder between the Cr and Al atoms, the occupancies have been refined in such a way that the partial occupancy of Cr on Al site and Al on Cr site fulfil the condition that the total occupancy of atoms is conserved on both the sites. However, since the neutron scattering amplitudes of Cr (3.6~fm) and Al (3.5~fm) are almost identical, no conclusive result could be obtained about the Y--Z type disorder \cite{SvyazhinJETP13}. On the other hand, the neutron scattering cross section is different for the Co (2.5~fm) and Cr (3.6~fm). Therefore, next we have introduced a disorder between Co and Cr atoms; however, no improvement in the refinement parameters was achieved. This suggests the absence of a disorder between Co and Cr for the $x=$ 0 sample, which is expected as the presence of the (200) reflection in X-ray/neutron diffraction patterns rule out this possibility. Neutron diffraction of Co$_2$MnSi Heusler alloy shows no antisite disorder between Mn and Si; however about 10\% antisite disorder was found between Co and Mn in arc-melted samples \cite{RaphaelPRB02}.

The neutron powder diffraction pattern taken at 170~K, i.e. at a temperature well within the magnetically ordered state, does not show additional Bragg reflections [see Fig.~\ref{fig2}(a)]. This rules out the presence of an antiferromagnetic spin structure and points towards a ferro or ferrimagnetic structure \cite{MukadamPRB16,RameshJPCM12}. In the present case we observe only a slight increase in intensity of the Bragg peaks at 170~K as compare to 400~K. This makes it difficult to refine the diffraction pattern measured at 170~K with a magnetic phase and extract useful information about the magnetic structure \cite{LazpitaNJP13}. A simulation with the magnetic moment determined by magnetization measurements did indicate that the slight increase in the Bragg peak intensities is in agreement with expected ferromagnetic structure. However, the change in the Bragg peak intensity due to thermal effects, i.e. the Debye-Waller factor would give a similar change. Therefore a more detailed analysis of the data is required. 
\begin{figure}[h]
\includegraphics[width=3.0in]{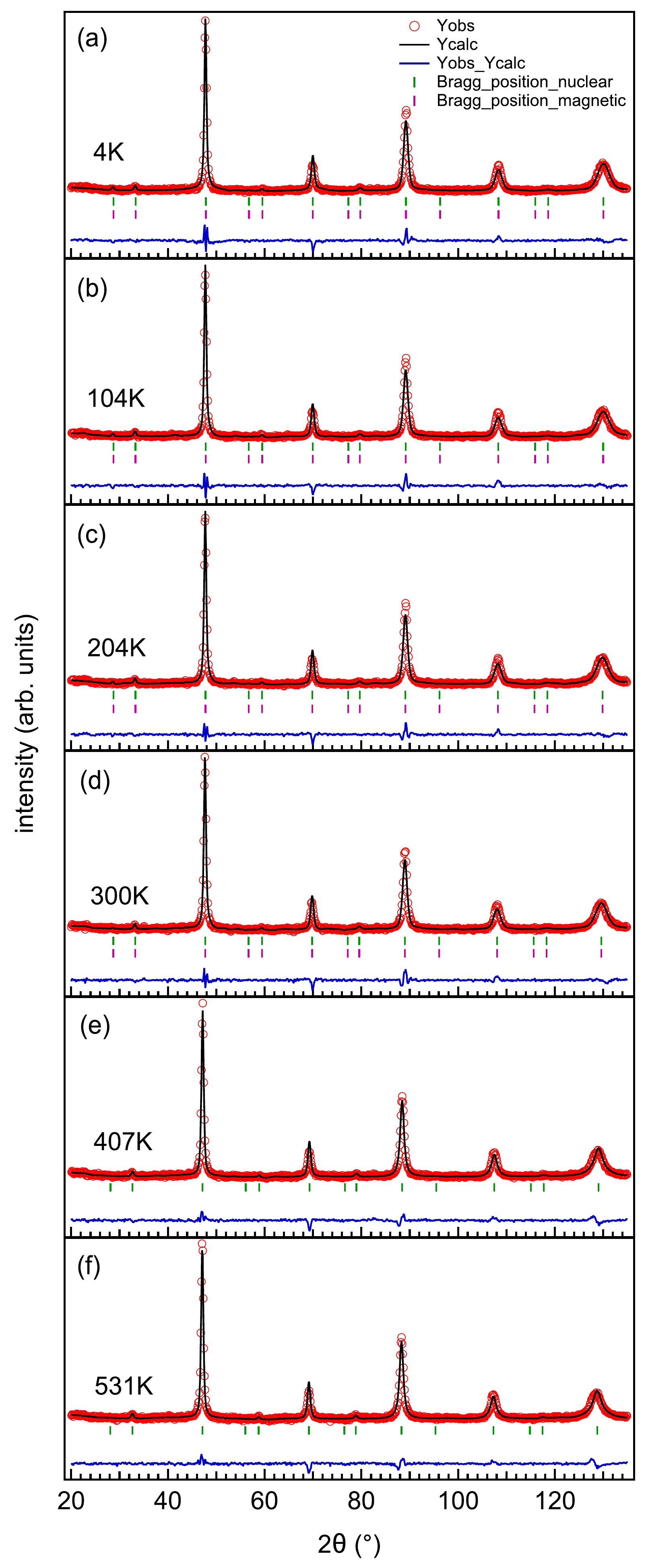}
\caption{The Rietveld refined neutron diffraction data at selected temperatures of the $x$= 0 sample measured at the high intensity defrectometer WOMBAT.}
\label{fig4}
\end{figure}
In order to detect distinct changes in the diffraction pattern across the magnetic phase transition temperature, we have used the high intensity diffractometer WOMBAT to record the data in a large temperature range across the magnetic phase transition. Figure~\ref{fig3} shows the neutron diffraction patterns of the $x=$ 0 sample measured at temperatures from 4~K to 531~K. At the first glance no obvious change is present in the diffraction pattern apart from the expected shift of the Bragg peaks towards lower 2$\theta$ value with increasing sample temperature, which can be attributed to the thermal expansion of the material. 

\begin{figure}
\includegraphics[width=3.3in]{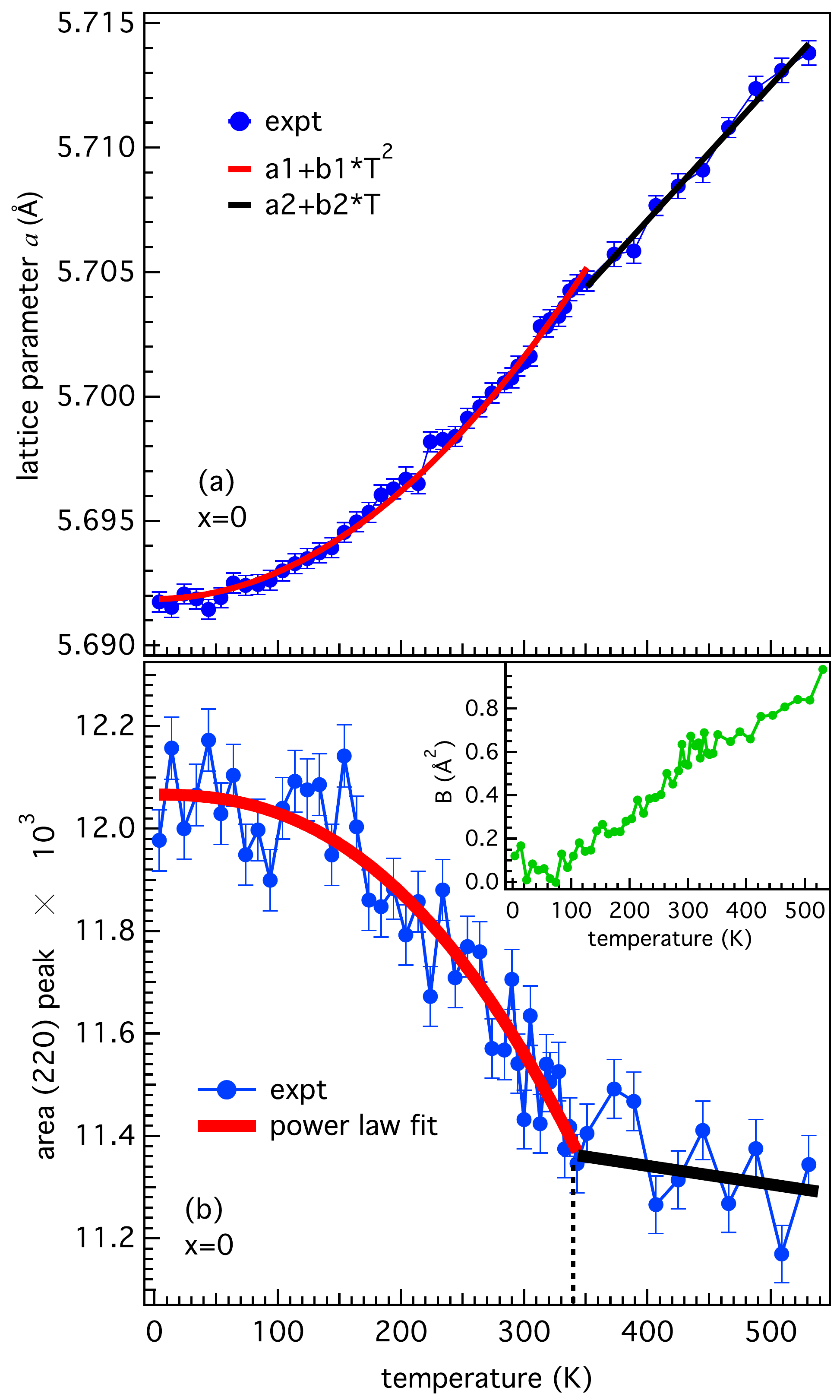}
\caption{(a) Variation in lattice constant $a$ and (b) corrected intensity of the (220) peak with temperature for the $x=$ 0 sample from the neutron diffraction data measured at the high intensity instrument WOMBAT. Inset in (b) shows the calculated $B$ parameter with sample temperature.}
\label{fig5}
\end{figure}

We have performed Rietveld refinements using the F{\it m$\bar{3}$m} space group for all temperatures, as shown in Fig.~\ref{fig4} for selected temperatures. The magnetization measurements show an increase in the magnetic moment below T$_{\rm C}=$ 330~K [see Figs.~\ref{fig1}(d--f)]. The intensity of nuclear Bragg reflections should increase below T$_{\rm C}$ in case of ferro- or ferri-magnetic ordering. Therefore, we have included a ferromagnetic structure in the Rietveld refinements of the diffraction pattern measured below the transition temperature. The refinement parameters for the pattern taken at 4~K are $\chi^2=$ 3.5, Bragg R-factor = 2.45, RF-factor = 1.68 and magnetic R-factor = 8.21, which confirms the high quality of the fits \cite{Ravel02}. The obtained value of total magnetic moment is 1.35$\pm$0.05~$\mu_{\rm B}$, which is close to the value of 1.45~$\mu_{\rm B}$ observed in M--H magnetization measurements, see Fig.~\ref{fig1}(g), but is significantly lower than expected from the Slater-Pauling rule. This would motivate for future investigation of these alloys using X-ray magnetic circular dichroism to understand the complex magnetic interaction and any possibility of orbital contribution. The neutron diffraction data at each temperature have been fitted using Rietveld refinement 
and the obtained lattice parameter $a$ is plotted as a function of sample temperature in Fig.~\ref{fig5}(a). Further analysis reveals that the lattice constant increases with the square power of the temperature below $\approx$350~K while exhibiting an essentially linear behavior for higher temperatures. The change in the lattice parameter $a$ follows the expected behavior for thermal expansion, which follows the Bose-Einstein statistics. The obtained constants from the fitting are: a1 = 5.6919~\AA, b1 = 1.08~\AA/K$^2$, a2 = 5.6854~\AA, and b2 = 5.4192$\times$10$^{-5}$~\AA/K. The calculated value of thermal coefficient is found to be 0.989$\times$10$^{-5}$/K, which is consistent with ref.~\cite{YanJPED08}. 

In order to determine the magnetic contribution of the Bragg peak intensity, we have plotted the total area under the most intense (220) reflection peak (I$_{\rm 220}$) with temperature (not shown). Upon decreasing temperature the intensity I$_{\rm 220}$ is fairly constant down to the magnetic phase transition and then increases as the sample goes from the paramagnetic to the ferromagnetic state. 
This behavior is consistent with the temperature dependence of the magnetization [Fig.~\ref{fig1}(d)] and 
with the reported behavior on similar Heusler alloys \cite{MukadamPRB16}. 
Note that the moment of the $x=$ 0 sample is quite weak and the magnetic phase transition takes place at a relatively high temperature where thermal effects can influence the intensity of the Bragg peaks as well. Therefore, it is important to determine the effect of temperature on the structural Bragg peaks, 
which is defined by the Debye-Waller parameter $B$. The magnetic structure factor decreases with 2$\theta$ and the Bragg peak at highest 2$\theta$ (=130$\degree$) has no significant magnetic contribution and its temperature dependence can be attributed to thermal effect only. In order to calculate the $B$ parameter, the intensity of (620) Bragg peak (2$\theta$=130$\degree$) at a particular temperature is expressed as \cite{InagakiJMS71}: $I_{620} (K)= I_0 \times exp (-2B sin^2\theta/\lambda^2)$, where $\theta$ is the diffraction angle, $\lambda$ is the wavelength of the X-ray/neutron source and $I_0$ is the intensity when there are negligible thermal vibrations in the atoms at low temperatures (4~K in the present case). We assumed that the value of $B$ is 0.05 at 4~K, which is reasonable as $B$ should be zero at 0~K. The calculated values of $B$ (using above equation) parameter, as shown in the inset of Fig.~\ref{fig5}(b) as a function of sample temperature. These values of $B$ were used to calculate corrected intensity of the (220) Bragg peak at each temperature are shown in Fig.~\ref{fig5}(b). The intensity increases below the magnetic phase transition temperature 330~K and follows the increase in magnetic moment where the magnetic neutron scattering intensity is proportional to M$^2$ \cite{TaheriPRB16, ReehuisPRB12}. 

\begin{figure*}
\includegraphics[width=7.1in]{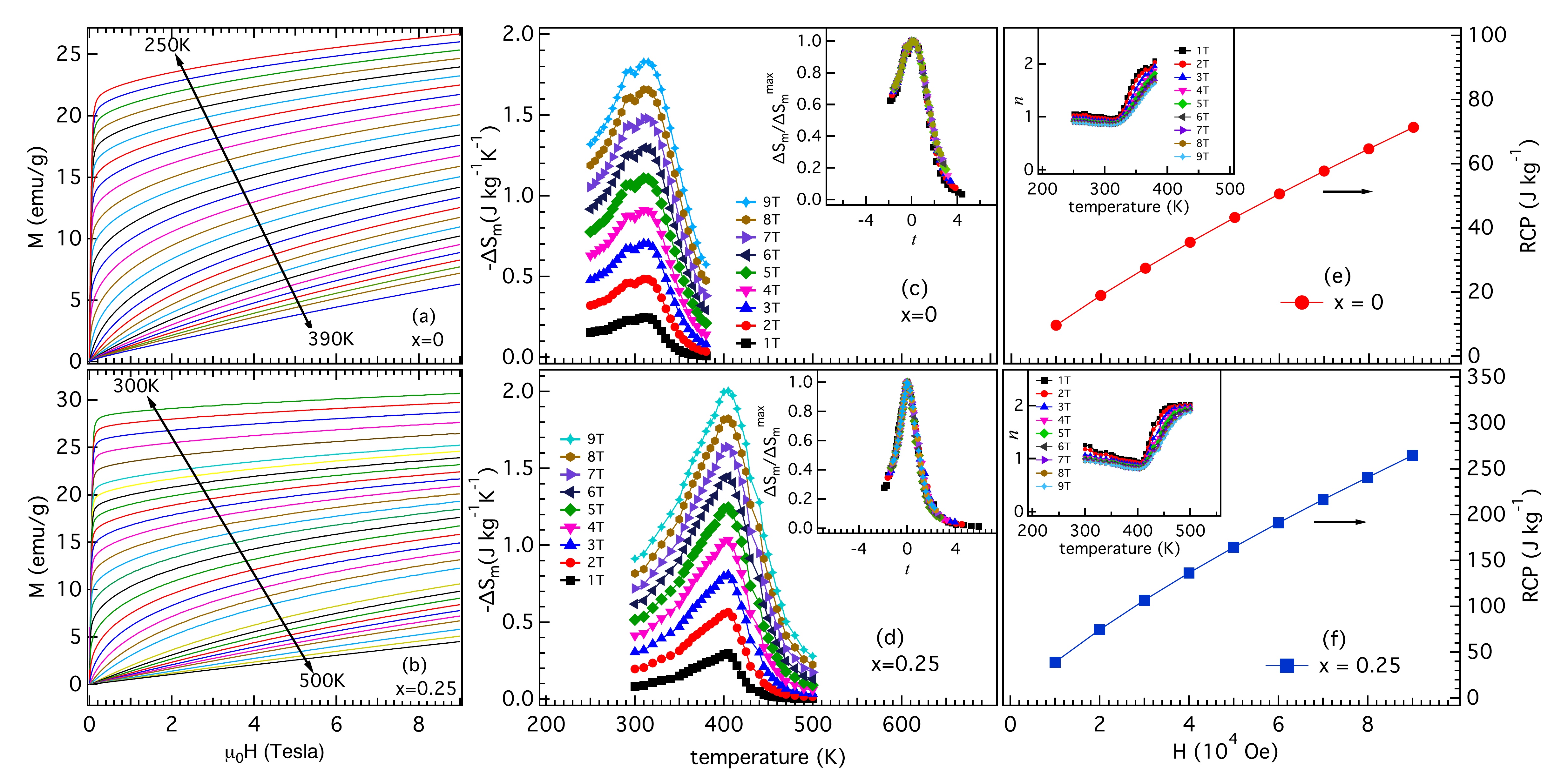}
\caption{(a, b) Isothermal magnetization measured at various temperatures across the Curie temperature, (c, d) change in the magnetic entropy $\Delta$S$_m$ with temperature, (e, f) variation of the relative cooling power (RCP) with magnetic field, for the Co$_2$Cr$_{1-x}$Ti$_x$Al samples with $x=$ 0, and 0.25. The insets in (c) and (d) are the normalized change in entropy with scaled temperature ($t$). The insets in (e) and (f) show the variation of exponent $n$ with temperature at various magnetic fields.} 
\label{fig6}
\end{figure*} 

In order to study the magnetocaloric effect (MCE) in these materials we have measured the isothermal magnetization for the Co$_2$Cr$_{1-x}$Ti$_x$Al ($x=$ 0, 0.25) samples in the vicinity of their respective magnetic phase transition temperature T$_{\rm C}$ [see Figs.~\ref{fig6}(a, b)]. The Maxwell's equation: $\Delta S_m= \int_{0}^{H} [\delta M(H,T)/\delta T]_H dH$ was used to extract  change in the magnetic entropy ($\Delta S_m$) from these isothermal magnetization curves. Figures~\ref{fig6}(c, d) show $\Delta S_m$ versus sample temperature in the vicinity of T$_{\rm C}$ for magnetic fields from 1~Tesla to 9~Tesla. For the $x=$ 0 and 0.25 samples the temperature corresponding to the maximum in the entropy $\Delta S_m^{max}$ (T$_{pk}$) is close to T$_{\rm C}$ as observed in Figs.~\ref{fig1}(d, e). Moreover, the measured $\mid\Delta S_m\mid$ value is about 2.0 J kg$^{-1}$ K$^{-1}$ at 9~Tesla for the $x=$0.25 sample. A comparable value of $\mid\Delta S_m\mid$ = 2.55~J kg$^{-1}$ K$^{-1}$ was reported for the Co$_2$CrAl sample \cite{PandaJALCOM15}. It is important to determine the precise magnetic interactions in Co$_2$Cr$_{1-x}$Ti$_x$Al samples. Therefore, we have used the MCE curves for further analysis. First to examine the field dependence of the experimental $|\Delta S_m|$ data at the different magnetic fields, the value of a local exponent $n$ can be calculated as $n= d(ln |\Delta S_m|)/d(lnH)$ \cite{ShenJAP02}, which depends on magnetic field and temperature. In the insets of Figs.~\ref{fig6}(e, f), we show the plot of exponent $n$ across the magnetic phase transition for the $x=$ 0 and 0.25 samples, respectively. Note that below T$_{\rm C}$ the value of $n$ is $\approx$1, which decreases slightly at T$_{\rm C}$ and then increases up to 2 above T$_{\rm C}$ in the paramagnetic region \cite{LawNatureCom18}. However, we have observed that the value of $n$ does not reach 2/3 at T$_{\rm C}$, as predicted if the system follow only the mean field model. This suggests deviation from the mean field type interactions in these samples. Moreover, various theoretical studies associate $\Delta S_m(T)$ as a function of scaled temperature with a second order phase transition for a ferromagnetic system \cite{FrancoAPL2006,FrancoEPL2007,PhanPRB2011,FrancoJMMM2009,MarcelaPRB2010,FrancoJPCM2008}. Here the scaled temperature, $t$ is defined as follows \cite{FrancoAPL2006, Dong08}: %
\begin{equation}
\begin{aligned}
t = 
\begin{cases}
-(T-T_{pk})/ (T_{r1}-T_{pk}); 	\quad T\leq{T_c}\\
(T-T_{pk})/ (T_{r2}-T_{pk}); 	\quad T>{T_c}
\end{cases}
\end{aligned}
\label{eq2}
\end{equation}
where $T_{r1}$ and $T_{r2}$ are the reference temperatures below and above the magnetic phase transition temperature, respectively, which correspond to certain $f$ ($=\Delta S_m(T)$/$\Delta S_{m}^{max}$) values i.e. $\Delta S_m(T)$/$\Delta S_{m}^{max}$ versus $t$. $T_{pk}$ is the temperature related to the maximum of $\Delta S_{m}$. The temperatures $T_{r1}$ and $T_{r2}$ have been calculated using $f=$ 0.8 and 0.5 for the $x=$ 0 and 0.25 samples, respectively. It has been demonstrated that $\Delta S_m(T)$/$\Delta S_{m}^{max}$ versus scaled temperature for different applied magnetic fields overlap into one single curve \cite{MarcelaPRB2010, FrancoJPCM2008}, as shown in the insets of Figs.~\ref{fig6}(c, d) for the $x=$ 0 and 0.25 samples. This is a universal behavior of the MCE for ferromagnetic materials with a second order phase transition. 

The relative cooling power (RCP) is a further important parameter for the suitability of these alloys as magnetocaloric materials, where a high value is favorable for refrigeration applications. The Co$_2$Cr$_{1-x}$Ti$_x$Al Heusler alloys have the benefit to possess high RCP values due to their broad second order phase transition at high temperatures. The RCP can be calculated by taking the product of the maximum entropy change and the temperature range of the full width half maximum, i.e. ${\rm RCP}= |-\Delta S_m|\times \delta T_{FWHM}$, where $\delta T_{FWHM}$ is determined by fitting the magnetic entropy curves with a Gaussian function. The RCP value is about 74.5~J kg$^{-1}$ at 1~Tesla and increases to about 264.5~J kg$^{-1}$ at 9~Tesla magnetic field for the $x=$ 0.25 sample, as shown in Fig.~\ref{fig6}(f). This is significantly higher than the value for the $x=$ 0 sample, see Figs.~\ref{fig6}(e, f). Therefore, the doped Co$_2$ based Heusler alloys are potential candidates for MCE applications as they show a second order magnetic phase transition, their T$_{\rm C}$ can be tuned systematically to the desired temperature range and the relative cooling power can be improved significantly through Ti doping \cite{PandaJALCOM15, PriyankaXiv19}.  

\begin{figure}
\includegraphics[width=3.4in]{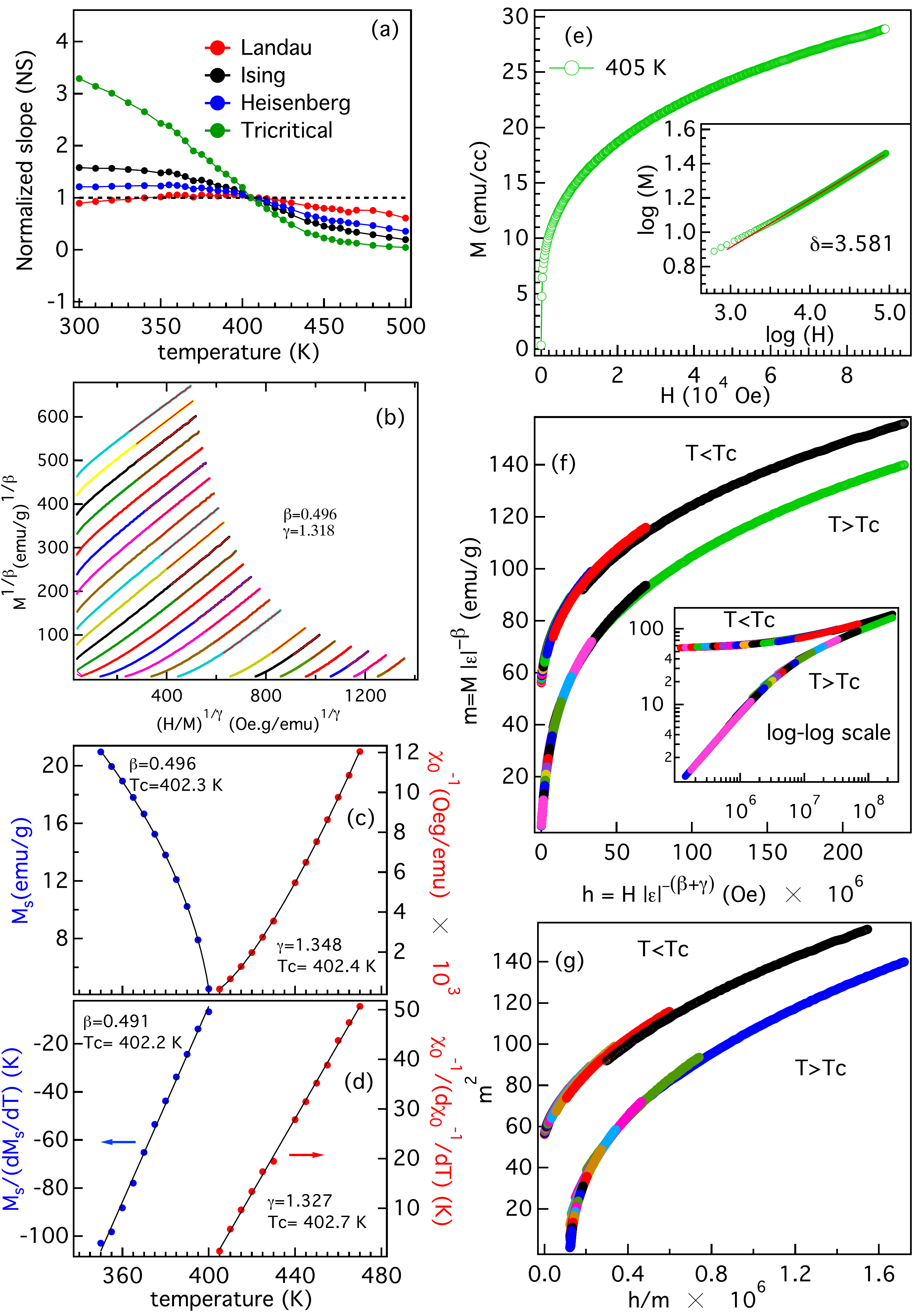}
\caption{Critical behavior analysis of Co$_2$Cr$_{0.75}$Ti$_{0.25}$Al sample: (a) The temperature dependence of normalized slope (NS) of four different models, (b) modified Arrot plot $M^{1/\beta}$ vs. $(H/M)^{1/\gamma}$ of the isotherms, (c) the plot of M$_{\rm S}$ (left axis) and $\chi_{0}^{-1}$ (right axis) and (d) the Kouvel-Fisher plot of M$_{\rm S}$ (left axis) and $\chi_{0}^{-1}$ (right axis) as a function of temperature. (e) the M vs H curve along with the log-log scale plot in the inset. (f) the renormalized magnetization ($m$) plotted as a function of renormalized field $h$ and (g) the plot in the form of $m^2$ vs $h/m$.}
\label{fig7}
\end{figure} 

For a further understanding of the magnetic interactions in these Heusler alloys, we did chose the $x=$ 0.25 sample due to its relatively sharp magnetic phase transition and performed a critical behavior analysis in detail. For this purpose, the critical exponents $\beta$ and $\gamma$ can be determined using the following equations:
\begin{equation}
M_S(T) = M_0(-\epsilon)^\beta; \quad \epsilon <0, T<T_c
\label{eq4}
\end{equation}
\begin{equation}
\chi_0^{-1} = (h_0/M_0)(\epsilon)^\gamma; \quad \epsilon >0, T>T_c
\label{eq5}
\end{equation}
where $\epsilon$= [(T-T$_{\rm C}$)/T$_{\rm C}$] is the reduced temperature, $h_0/M_0$ is the critical amplitude, and the exponents $\beta$ and $\gamma$ are related to the saturation magnetization M$_S$(T) and the inverse initial magnetic susceptibility $\chi_0^{-1}$ \cite{Arrott67}. The values of the critical exponents can give an idea about the different types of magnetic interactions present in the system. Therefore we have used different theoretical models for modified Arrot plots, where we have applied different initial values for $\beta$ and $\gamma$ for the mean field model (0.5, 1), the 3D Heisenberg model (0.365, 1.386), the Ising model (0.325, 1.24) and the tricritical model (0.25, 1). The corresponding M$^{1/\beta}$ was plotted versus (H/M)$^{1/\gamma}$ based on the Arrott-Noakes equation of state \cite{Arrott67}, for the $x=$ 0.25 sample (data not shown here). For the best fit model, the plot should have straight and parallel lines in the high magnetic field region. We have fitted the curves in the high field region with the straight line and the line which passes through origin corresponds to the T$_{\rm C}$ value. Moreover, in order to determine best fit model, we have defined the normalized slope as $S$($T$)/$S$($T_{\rm C}$), where $S(T)$ is the slope of M$^{1/\beta}$ vs. (H/M)$^{1/\gamma}$ at the temperature T and $S(T_c)$ is the slope at T$_{\rm C}$. Fig.~\ref{fig7}(a) shows the normalized slope as a function of temperature for the $x=$ 0.25 sample, fitted for the modified Arrott plots for the different models. The value of the normalized slope found to be close to one, which indicates the mean field model as the best fit model. The calculation of exact value of the critical exponents is an iterative method \cite{Zhang12}, where we use equations \ref{eq4} and \ref{eq5} to obtain M$_S$ (T, 0) and $\chi_0^{-1}$(T, 0) at zero field from the linear extrapolation of a straight line fit of the high field region. We start with the initial values of $\beta$=0.5 and $\gamma$=1 for the mean field model and fitted the curves of M$^{1/\beta}$ vs (H/M)$^{1/\gamma}$ using equations \ref{eq4} and \ref{eq5}, which gives stable values for the critical exponents i.e. $\beta=0.496$, $\gamma=1.318$ for the $x=$0.25 sample, as shown in Fig.~\ref{fig7}(b). Now we use these values of $\beta$ and $\gamma$ and the fit shows straight lines in the high field region. However, the lines are not perfectly straight probably due to a complex domain structure in the low field region. Furthermore, we have plotted M$_S$ and $\chi_0^{-1}$ as a function of temperature from Fig.~\ref{fig7}(b) by taking the slope values on M$^{1/\beta}$ and (H/M)$^{1/\gamma}$, respectively. In Fig.~\ref{fig7}(c), the fit of a power-law function following equations \ref{eq4} and \ref{eq5} gives the values of $\beta=0.496$ and $\gamma=1.348$ for the $x=$ 0.25 sample, which are close to the one obtained from Fig.~\ref{fig7}(b). This confirms the reliability of the parameters $\beta$ and $\gamma$.

It is important to further verify the accuracy of these critical exponents obtained from the modified Arrott plot. Therefore, equations \ref{eq4} and \ref{eq5} have been expressed following the Kouvel-Fisher method, as below:
\begin{equation}
M_S(T)/[dM_S(T)/dT] = \beta^{-1} (T-T_c)
\label{eq7}
\end{equation}
\begin{equation}
\chi_0^{-1}(T)/[d\chi_0^{-1}(T)/dT] = \gamma^{-1} (T-T_c)
\label{eq8}
\end{equation}
\begin{equation}
{\rm log} M(H,T_c) = \delta^{-1} {\rm log} H 
\label{eq9}
\end{equation}
where $\delta$ corresponds to the critical magnetization at T$_{\rm C}$. The critical exponents $\beta$ and $\gamma$ have been calculated from equations \ref{eq7} and \ref{eq8}, respectively. The slope of straight line fit to the plots of $M_S$ (d$M_S/dT)^{-1}$ and $\chi_0^{-1}$ (d$\chi_0^{-1}/dT)^{-1}$ as a function of temperature, as shown in Fig.~\ref{fig7}(d), gives the values of $1/\beta$ and $1/\gamma$. The value of $\delta$ is obtained using equation \ref{eq9} where the slope of a straight line fit of log(M) vs. log(H) curve at T$_{\rm C}$ gives 1/$\delta$ [see Fig.~\ref{fig7}(e)]. Using equations \ref{eq7}, \ref{eq8} and \ref{eq9}, the obtained values are $\beta$=0.491, $\gamma$=1.327, and $\delta$=3.581 for the $x=$ 0.25 sample. These values are quite close to reported recently for similar Heusler compound Co$_2$TiGe by Roy {\it et al.} in ref.~\cite{RoyPRB19} and are comparable to the theoretical values \cite{KaulJMMM85}. Furthermore, we have used the Widom scaling relation $\gamma - \beta (\delta -1) = 0$ in order to validate the accuracy of $\delta$ value, which was found to be in agreement with the one obtained from the critical isotherm, the Kouvel-Fisher method and the modified Arrott plot.
 
The obtained values of $\beta$ and $\gamma$ suggest two types of magnetic interactions in the system. Below T$_{\rm C}$ the estimated $\beta$ value is close to mean field model, which indicates long range magnetic interactions, while the value of $\gamma$ deviates from mean field towards 3D Heisenberg model above T$_{\rm C}$, which suggests the presence of the local magnetic interactions. This behavior is in agreement with a recent study on Co$_2$TiGe \cite{RoyPRB19} as well as other inter-metallic compounds \cite{HalderJAP11}, where the critical exponents suggest the presence of complex magnetic interactions. Therefore, the values of the critical exponents offers the opportunity to check the scaling equation of the magnetic state, which can be performed as follows: 
\begin{equation}
M(H,\epsilon) = \epsilon^{\beta}f_{\pm} (\frac{H}{\epsilon^{\beta+\gamma}})
\label{eq3}
\end{equation}
where $f_-$ and $f_+$ are regular analytical functions below and above T$_{\rm C}$, respectively. This scaling equation of state describes the relationship between M (H,$\epsilon$), H and T$_{\rm C}$. Following the equation~(\ref{eq3}) the scaled magnetization $m=\epsilon^{-\beta} M(H, \epsilon$) as a function of scaled field $h=\epsilon^{-(\beta+\gamma)}H$ should show two distinct branches across the Curie temperature T$_{\rm C}$. Therefore, in Fig.~\ref{fig7}(f), we show the scaled value of $m$ versus the scaled value of $h$. This clearly demonstrates the expected behavior and is a further validation of the above determined values of critical exponents ($\beta$, $\gamma$) and T$_{\rm C}$. The same plot in log-log scale is shown in the inset of Fig.~\ref{fig7}(f). The splitting of $m^2$ vs $h/m$ into two branches across the T$_{\rm C}$ also indicates the reliability of the determined exponents and T$_{\rm C}$ as depicted in Fig.\ref{fig7}(g).

Notably the values of the critical exponents obtained by the detailed analysis for the $x=$ 0.25 sample do not follow any conventional universality class and predict that the system deviates from the mean field towards the 3D Heisenberg model i.e. long-range type as well as local magnetic interactions below and above T$_{\rm C}$, respectively. This indicates that the nature and range of complex spin interactions are playing a crucial role in determining the magnetic properties of this compound. It has been reported that isotropic exchange interactions between spins can give rise to extended spin interactions \cite{FisherPRL72, RoyPRB19, PriyankaXiv19}, which we can represent in terms of the exchange distance $J(r)$. In case of long range exchange interactions $J(r)$ decays as $J(r)\approx r^{-(D+\sigma)}$; whereas short range interactions decays according to $J(r)\approx e^{-r/b}$ where $r$ is the distance, $D$ is the lattice dimensionality, and $b$ is a specific scaling factor \cite{FisherPRL72}. Here, the crucial parameter is $\sigma$, which determines the interaction range and depending on its value, i.e. if $\sigma<$1.5 or $>$2 indicates the presence of long range or short range spin interactions, respectively. Therefore, we have calculated the $\sigma$ value following the renormalization group approach where the susceptibility exponent $\gamma$ can be expressed by the equation below \cite{FisherPRL72, Fisher74}: 
\begin{equation}
\begin{split}
\gamma = 1+ \frac{4}{D}\frac{d+2}{d+8} \Delta \sigma + \frac{8(d+2)(d-4)}{D^2(d+8)^2}\\
\times\bigg[1+\frac{2G(\frac{D}{2})(7d+20)} {(d-4)(d+8)}\bigg]\Delta \sigma^2
\label{eq12}
\end{split}
\end{equation}
where $d$ is the spin dimensionality, and $\Delta \sigma$ and $G(\frac{D}{2})$ can be defined as $\sigma-\frac{D}{2}$ and $3-\frac{1}{4}(\frac{D}{2})^2$, respectively. As discussed above, the value of $\sigma$ can give an idea of the length scale of the spin interactions. Therefore, $\sigma$ is calculated using above equation and the experimental value of $\gamma=$ 1.348. By considering $\{D:d\}$=$\{3:3\}$ we found a $\sigma$ value of about 1.85. We note here that for a value of $\sigma$ between 1.5 and 2, the system possesses extended type spin interactions other than the existing universality classes. With the obtained value of $\sigma$ the other critical exponents can be calculated using the expressions: $\upsilon=\gamma/\sigma$, $\alpha=2-\upsilon d$, $\beta=(2-\alpha-\gamma)/2$ and $\delta=1+\gamma/\beta$, where $\upsilon$ is the exponent of the correlation length. By taking the $\sigma=$ 1.85 the critical exponents are determined to be $\beta$=0.404 and $\delta$=4.337, which deviate from the experimental values (as discussed above for the $x=$ 0.25 sample) and further suggests the presence of complex magnetic interactions in the sample. Interestingly, our study indicates that the long-range exchange interactions dominate below T$_{\rm C}$, which are coupled with the 3D Heisenberg type short range spin interactions across the magnetic phase transition \cite{RoyPRB19}. This motivates for future investigations to understand its origin.  

\section{\noindent ~Conclusions}

The structural and magnetocaloric properties of Co$_2$Cr$_{1-x}$Ti$_x$Al Heusler alloys have been investigated using X-ray and neutron diffraction, and magnetization measurements. The Rietveld refinement of the X-ray diffraction data indicates an increase of the lattice parameter and unit cell volume with Ti substitution. We observed a second order phase transition from paramagnetic to ferromagnetic where the Curie temperature systematically increases from 330~K for the $x=$ 0 sample to 445~K for the $x=$ 0.5 sample. The neutron powder diffraction data of the $x=$ 0 sample taken  across the magnetic phase transition in a large temperature range confirm the structural stability. Further, the magnetic moment of the $x=$ 0 sample is found to be 1.45~$\mu_{\rm B}$/{\it f.u.} from magnetization measurements, which is in good agreement to the value of 1.35$\pm$0.05~$\mu_{\rm B}$/{\it f.u.} extracted by the Rietveld analysis of neutron powder diffraction pattern measured at 4~K. By analysing the temperature dependent neutron diffraction data for the $x=$ 0 sample, we find that the change in the magnetic intensity of the most intense Bragg peak (220) is consistent with the increase in the magnetic moment with temperature. Moreover, an enhancement in the change in the magnetic entropy $\Delta S_m$ and the relative cooling power was observed for the $x=$ 0.25 sample. Furthermore, our study reveals the presence of long-range ferromagnetic ordering below T$_{\rm C}$, which deviates towards short 3D Heisenberg type spin interactions above T$_{\rm C}$ based on the analysis of the critical behavior across the second order phase transition with the extracted exponents values of $\beta\approx$ 0.496, $\gamma\approx$ 1.348, and $\delta\approx$ 3.71 for the $x=$ 0.25 sample, which is further supported by the interaction range ($\sigma=$ 1.85) analysis. 

\section{\noindent ~Acknowledgments}

This work was financially supported by the BRNS through DAE Young Scientist Research Award to RSD with project sanction No. 34/20/12/2015/BRNS. PN and GDG thank the MHRD, India for fellowship through IIT Delhi. Authors acknowledge the Physics department, IIT Delhi and UGC-DAE CSR, Mumbai centre for X-ray diffraction and magnetic measurements. PN, GDG and RSD thank Sanjay Singh for help in neutron diffraction data analysis and useful discussion. We thank V. Siruguri for support in performing the magnetic measurements. RSD also gratefully acknowledges the financial support from the department of science \& technology (DST), India through Indo-Australia early and mid-career researchers (EMCR) fellowship, administered by INSA (sanction order no. IA/INDO-AUST/F-19/2017/1887) for performing the neutron measurements at ANSTO, Australia. CU thanks the Australian Research Council (ARC) for the support through the Discovery Grant DP160100545.

\end{document}